 \documentclass[final,5p,times,twocolumn,authoryear]{elsarticle}

\usepackage[colorlinks=true,linkcolor=blue,citecolor=blue,urlcolor=blue]{hyperref}

\usepackage{amssymb}
\usepackage{lipsum}
\usepackage{graphicx}
\usepackage{xcolor}
\usepackage{subcaption}
\usepackage{amsmath}
\usepackage{float}
\usepackage{lineno}

\journal{High Energy Astrophysics}

\begin{document}

\begin{frontmatter}



\title{Profile Reconstruction from Temporally Stable Emission Components for Timing PSR J1713+0747}


\author[RAC]{Shaswata Chowdhury\corref{cor1}}
\ead{shaswata.phyres@gmail.com}
\cortext[cor1]{Corresponding author}
\author[NCRA]{M. A. Krishnakumar}
\author[IITH]{Sharika Dhakappa}
\author[TIFR]{Vidit Singh}
\author[NWU,NITheCS]{Debabrata Deb}
\author[INAF]{Jyotijwal Debnath}
\author[IISERB]{Kaustubh Rai}
\author[INAF]{Pratik Tarafdar}
\author[IISERTVM]{Abhimanyu Susobhanan}
\author[PRL]{Churchil Dwivedi}
\author[NCRA,IITR]{Bhal Chandra Joshi}
\author[IITH]{Shantanu Desai}
\author[IITD]{Neelam Dhanda Batra}
\author[UCT]{Jaikhomba Singha}
\author[NCRA]{Himanshu Grover}
\author[IMSc,HBNI]{Manjari Bagchi}
\author[IISERB]{Mayuresh Surnis}
\author[IISc]{Avinash Kumar Paladi}
\author[IITH]{Aman Srivastava}
\author[RRI,Christ]{Arul Pandian B.}
\author[IBS,UW]{Suruj Jyoti Das}
\author[IITI]{Jibin Jose}
\author[CEBS]{Kuldeep Meena}
\author[IITB]{Sushovan Mondal}
\author[Kumamoto]{K Nobleson}

\author[Kumamoto]{Keitaro Takahashi}
\author[IISERB]{Hemanga Tahbildar}
\author[IISERB]{Kunjal Vara}
\author[IISc]{Zenia Zuraiq}

\address[RAC]{Radio Astronomy Centre, National Centre for Radio Astrophysics, Tata Institute of Fundamental Research, Udhagamandalam, 643001, India}
\address[NCRA]{National Centre for Radio Astrophysics, SP Pune University Campus, Pune, Maharashtra, 411007, India}
\address[IITH]{Department of Physics, IIT Hyderabad, Kandi, Telangana 502284, India}
\address[TIFR]{Department of Astronomy and Astrophysics, Tata Institute of Fundamental Research, Mumbai 400005, Maharashtra, India}
\address[NWU]{Centre for Space Research, North-West University, Private Bag X6001, Potchefstroom 2520, South Africa}
\address[NITheCS]{National Institute for Theoretical and Computational Sciences (NITheCS), Stellenbosch 7604, South Africa}
\address[INAF]{INAF - Osservatorio Astronomico di Cagliari, via della Scienza 5, 09047 Selargius (CA), Italy}
\address[IISERB]{Department of Physics, IISER Bhopal, Bhauri Bypass Road, Bhopal, 462066, India}
\address[IISERTVM]{School of Physics, Indian Institute of Science Education and Research Thiruvananthapuram, Maruthamala P.O., Thiruvananthapuram 695551, Kerala, India}
\address[PRL]{Astronomy and Astrophysics Division, Physical Research Laboratory, Thaltej Campus, Thaltej, Ahmedabad 380059, Ahmedabad, Gujarat, India}
\address[IITR]{Department of Physics, Indian Institute of Technology Roorkee, 
Roorkee, Uttarakhand, 247667, India }
\address[IITD]{Department of Physics, Indian Institute of Technology Delhi, Hauz Khas, New Delhi 110016}
\address[UCT]{High Energy Physics, Cosmology and Astrophysics Theory Group (HEPCAT), Department of Mathematics and Applied Mathematics, University of Cape Town, Cape Town 7700, South Africa}
\address[IMSc]{The Institute of Mathematical Sciences, C. I. T. Campus, Taramani, Chennai 600113, India}
\address[HBNI]{Homi Bhabha National Institute, Training School Complex, Anushakti Nagar, Mumbai 400094, India}
\address[IISc]{Joint Astronomy Programme, Department of Physics, Indian Institute of Science, Bengaluru, Karnataka, 560012, India.}
\address[RRI]{Raman Research Institute, Bengaluru - 560080, India}
\address[Christ]{Department of Physics and Electronics, CHRIST (Deemed to be University), Bengaluru - 560029, India}
\address[IBS]{Particle Theory  and Cosmology Group, Center for Theoretical Physics of the Universe,
Institute for Basic Science (IBS),
 Daejeon, 34126, Korea}
\address[UW]{Faculty of Physics, University of Warsaw, Pasteura 5, 02-093 Warsaw, Poland}
\address[IITI]{Department of Astronomy, Astrophysics, and Space Engineering, Indian Institute of Technology Indore, Indore 453552, India}
\address[CEBS]{UM-DAE Centre for Excellence in Basic Sciences,  University of Mumbai, Vidyanagari, Mumbai 400098, India}
\address[IITB]{Department of Physics, Indian Institute of Technology Bombay, Mumbai 400076, India}
\address[Kumamoto]{Faculty of Advanced Science and Technology, Kumamoto University, 2-39-1 Kurokami, Kumamoto 860-8555, Japan}

\begin{abstract}
The assumption of long-term pulse-profile stability underpins high-precision pulsar timing and forms the basis of pulsar timing array experiments. However, several millisecond pulsars exhibit temporal profile variability that can introduce systematic biases in pulse time of arrival measurements and compromise timing precision. We present a profile-domain analysis of PSR J1713+0747 at low radio frequencies, in the 300–500 MHz band, using upgraded GMRT observations for the Indian Pulsar Timing Array experiment. We model frequency-resolved pulse profiles using a Bayesian Gaussian decomposition framework in which individual Gaussian components are associated with persistent emission regions through informative phase priors that permit modest temporal variations. By tracking the evolution of the decomposed components across observing epochs and frequency sub-bands, we identify central Gaussian components that remain precisely localized despite changes in the integrated pulse morphology. We then reconstruct pulse profiles with realistic noise using these central components and perform timing analysis. Our approach provides a physically motivated framework for mitigating pulse-profile variability and offers a generic methodology for recovering robust timing information from pulsars exhibiting profile evolution.
\end{abstract}




\end{frontmatter}




\section{Introduction}
\label{introduction}

Millisecond pulsars (MSPs) are among the most stable natural
rotators known and constitute the foundation of pulsar timing
array (PTA) experiments aimed at detecting nanohertz
gravitational waves,
and probing a wide range of astrophysical and fundamental
physics phenomena. The remarkable timing precision achieved
for the best MSPs relies on the implicit assumption that the
average pulse profile remains stable over timescales spanning
years to decades, allowing pulse times of arrival (ToAs) to be
determined by cross-correlating the observed pulse profiles with a high signal-to-noise template \citep{TaylorTiming,LorimerKramerHandBook}.

Although this assumption has proven remarkably successful for
precision timing, an increasing number of pulsars are now known
to exhibit long-term pulse-profile variability \citep{JacobsonBell2026_NANOGrav, Keith}. Pulse-profile variability may arise from a variety of physical and observational processes, including intrinsic magnetospheric state changes and mode changing \citep{Lyne2010}, stochastic pulse-to-pulse variability such as pulse jitter \citep{Jitter1, Jitter2}, propagation through the interstellar or interplanetary media \citep{Lam2016,Lam2018}, or instrumental systematics. The observed frequency dependence of pulsar profiles is commonly interpreted within the framework of radius-to-frequency mapping and pulsar emission geometry \citep{MitraRankin2002}. Even relatively subtle profile variations can bias template
matching, introduce systematic timing offsets, and ultimately
limit PTA sensitivity \citep{Brook2016}.

Among the PTA pulsars, PSR~J1713+0747 is one of the most precisely timed pulsars and plays a key role in PTA gravitational-wave searches~\citep{Foster, NanoGrav_GW}. In April 2021, however, this pulsar underwent a
sudden pulse-profile change \citep{Jaikhomba2021} that has persisted for several years.
Subsequent observational studies established that the event is
broadband in nature, exhibits frequency-dependent behavior,
and is more consistent with an intrinsic magnetospheric
reconfiguration \citep{Jennings2024, Rami2025} than with propagation effects in the
interstellar medium
\citep{Lin2021}. The event poses a
significant challenge for conventional pulsar timing because
the evolving pulse morphology violates the assumption of a
stationary template profile.

Several approaches have recently been proposed to mitigate the
impact of profile variability on pulsar timing. Principal
component analyses~\citep{PCA1}, profile-domain timing techniques~\citep{Lentati1, Lentati2}, and
Gaussian-component modeling \citep{Nathan} have demonstrated that modeling
the pulse profile directly can substantially reduce timing
biases while retaining otherwise unusable observations. Recent work by \cite{Shania2026} has also shown that
Gaussian-component modelling provides an effective way to
maintain high-precision timing across the 2021 profile-change event of PSR~J1713+0747, using the high frequency observations ($\sim$ 850 MHz and L-band). 

In this work, we develop a physically-motivated Bayesian Gaussian decomposition approach, using low-frequency InPTA Band-3 (300--500 MHz) observations, where propagation effects are more pronounced, making it essential to distinguish intrinsic profile evolution from chromatic propagation delays for precision timing. Guided by
broadband polarization measurements, informative yet flexible
phase priors are assigned to Gaussian components associated
with distinct emission regions, allowing for modest
frequency-dependent as well as temporal evolution while preserving their physical
identity. The time evolution of the individual components is then
used to identify those emission regions that remain stable
throughout the profile-change event. Pulse profiles are
subsequently reconstructed using only these stable components,
while preserving realistic observational noise and the original
metadata. By performing the decomposition on
frequency-resolved profiles, the proposed methodology retains
the chromatic information required for reliable epoch-wise
dispersion measure estimation before carrying out conventional
timing analysis. Although motivated by the profile-change event in
PSR~J1713+0747, the methodology presented here is readily adaptable to other pulsars exhibiting abrupt or gradual profile evolution, provided suitable morphological or polarimetric information is available to construct component priors. 

This paper is organized as follows.
Section~\ref{sec:data} describes the observations and data processing.
Section~\ref{sec:motivation_5gaussian} motivates the Gaussian decomposition from the pulse
profile morphology, while Section~\ref{sec:bayesian} presents the Bayesian
decomposition framework and the construction of the
morphology-driven priors. The evolution of the Gaussian
components and the reconstruction of stable pulse profiles are
presented in Sections~\ref{sec:evolution_gauss} and~\ref{sec:reconstruction}, respectively. These
reconstructed profiles are subsequently used for dispersion
measure estimation and timing analysis in Sections~\ref{sec:dm_estimation} and~\ref{sec:timing}.
Finally, Section~\ref{sec:conclusion} summarizes the main results and discusses
their broader implications.

\section{Observations and Data Processing}
\label{sec:data}

PSR~J1713+0747 is routinely observed as part of the Indian Pulsar
Timing Array \cite[InPTA:][]{Joshi2022} using the upgraded Giant Metrewave Radio
Telescope \cite[uGMRT:][]{sak+1991,gak+2017}. The InPTA observing strategy, instrumental
configuration, and standard data-reduction procedures are described
in detail in \cite{Prerna2025} and \cite{Pratik2022}; here we
summarize only those aspects directly relevant to the present work.

In this analysis, we use only the Band-3 ($300$--$500$\,MHz)
observations from the InPTA, from observing Cycle 37 to Cycle 48, i.e., MJD ranging from 58773 - 60947 (17 October 2019 -- 29 September 2025). Each observation spans a
contiguous bandwidth of 200\,MHz, providing sufficiently wide
frequency coverage to capture the intrinsic frequency evolution of
the pulse profile while simultaneously enabling precise estimation of
the dispersion measure through the strong frequency dependence of the
dispersive delay. Although simultaneous Band-5 observations are also
available, the Band-3 data are better suited for profile-domain
decomposition because they are recorded with substantially higher
phase resolution ($N_{\rm bin}=128$, $256$, and $512$ across the
observing baseline), whereas the majority of the Band-5 observations
contain only 64 phase bins. The higher phase resolution permits a more faithful Bayesian Gaussian decomposition of the pulse profile and a more reliable characterization of the individual emission regions. 
Furthermore, the stronger influence of chromatic propagation effects at these frequencies provides a stringent test of whether our proposed method in this paper can successfully isolate intrinsic profile evolution while preserving the frequency-dependent information required for precision timing.

The starting point of our analysis is the calibrated PSRFITS~\citep{PSRFITS_Hotan} archives
produced by the standard InPTA processing pipeline PINTA \citep{AbhimanyuPINTA2021}. Following the
standard InPTA processing, the frequency channels
are reordered from low to high frequency, the data are integrated in
time by collapsing all sub-integrations, and the local-oscillator
timing correction for observations between MJD~59217 and 59424
\citep{Pratik2022} is applied before further analysis. The
archives are subsequently refolded and de-dispersed using the EPTA-DR2-full timing solution~\citep{EPTADR2}
after removing all epoch-dependent DMX parameters and replacing them
with a single fiducial dispersion
measure corresponding to the adopted template epoch
(MJD~59300). This provides a common timing reference
for all observations prior to the profile-domain analysis. Finally, the pulse profiles from all the epochs are rotated in pulse phase by 0.08 so that the main emission lies near the centre of the pulse window.

As a final quality-control step, observations exhibiting unusually high noise levels, corrupted data, or anomalous pulse-phase alignment (e.g., the main pulse displaced by several phase bins from its expected location after applying the common timing solution) are excluded from further analysis. The remaining archives constitute the input dataset for the
Bayesian Gaussian decomposition described in the remainder of the paper.

\section{Polarimetric Motivation for the Choice of Gaussian Components}
\label{sec:motivation_5gaussian}

In this work, the decomposition of frequency-integrated pulse profiles into Gaussian components is guided solely by the morphology of the total intensity profile. However, the integrated profile can obscure underlying emission structures, particularly when multiple emission regions overlap in phase. Polarimetric observations provide an additional and physically motivated means of identifying these structures, as changes in the linear and circular polarization modes and transitions in the polarization position angle (PA) are often associated with distinct emission regions within the pulsar magnetosphere.

\begin{figure}[!h]
\centering

\begin{subfigure}{0.9\columnwidth}
    \centering
    \includegraphics[width=\linewidth]{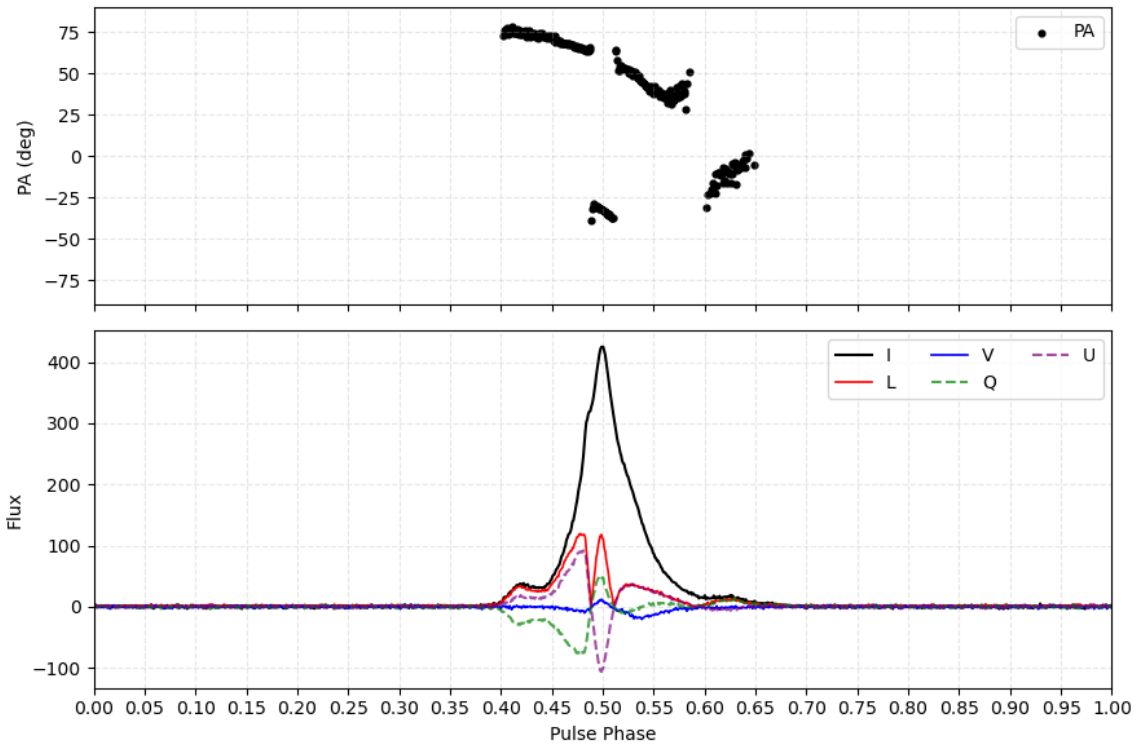}
    \caption{}
    \label{fig:1a}
\end{subfigure}

\vspace{0.2cm}

\begin{subfigure}{0.9\columnwidth}
    \centering
    \includegraphics[width=\linewidth]{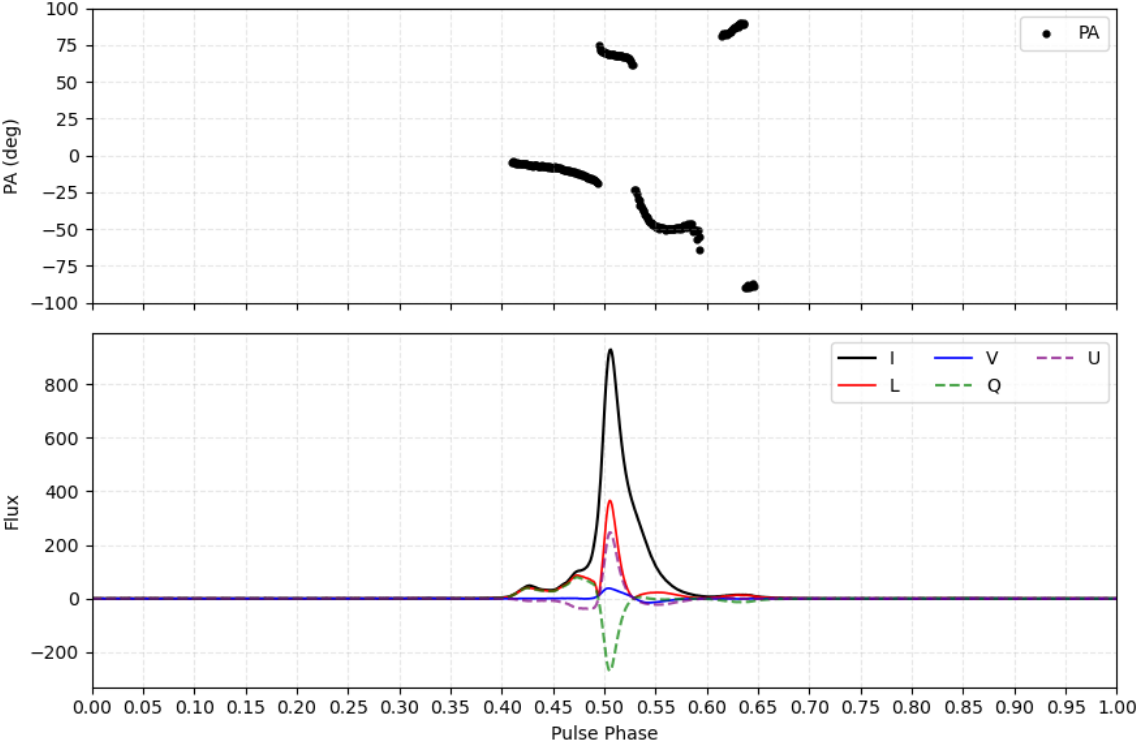}
    \caption{}
    \label{fig:1b}
\end{subfigure}

\caption{Full- polarisation pulse profiles of PSR J1713+0747 from the European Pulsar Network at two frequencies -- (a) 728 MHz and b) 1369 MHz observations with total intensity (I, black), linear polarisation (L, red), circular polarisation (V, blue) and the polarisation angle (PA, black).}
\label{fig:epn_profiles}
\end{figure}

PSR J1713+0747 has been observed in full polarization over a broad frequency range as available at the European Pulsar Network (EPN)\footnote{\url{https://psrweb.jb.man.ac.uk/epndb/}. See a list of contributions to this database at \url{https://psrweb.jb.man.ac.uk/epndb/about.html\#contrib}} database. Fig.~\ref{fig:epn_profiles} shows representative pre-event full-Stokes profiles at 728 and 1369 MHz, observed as part of the Parkes Pulsar Timing Array project~\citep{EPN_paper}. The linear polarization profile exhibits several distinct features accompanied by changes in the PA curve and transitions between polarization modes. Such behavior suggests the presence of multiple underlying emission regions contributing to the observed integrated profile.

Polarization provides a direct probe to the pulsar emission geometry, as the observed PA is determined by the orientation of the magnetic field lines with respect to the observer's line of sight. The observed pulse-profile morphology can evolve with observing frequency as a consequence of frequency-dependent emission geometry, including radius-to-frequency mapping~\citep{MitraRankin2002}. Nevertheless, the persistence of identifiable profile and polarization features across frequency, as seen for PSR J1713+0747~\citep{EPN_paper}, motivates associating these features with corresponding emission regions in the pulsar magnetosphere.

\begin{figure}
	\centering 
	\includegraphics[width=0.5\textwidth]{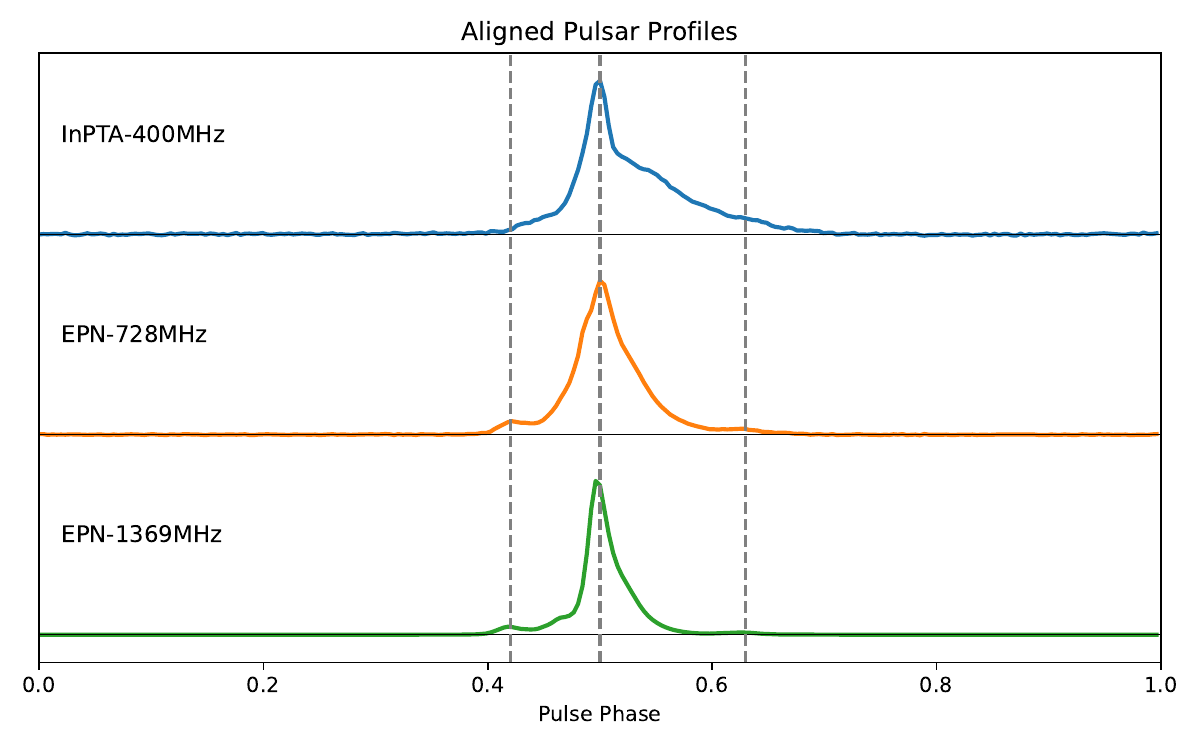}	
	\caption{Phase-aligned total intensity pulse profiles of PSR J1713+0747 across different frequencies. Profiles from InPTA Band-3 (400MHz, blue), EPN (728MHz and 1369 MHz) are overlaid after phase alignment. The three dashed lines represent the three features that remain consistent across these frequency bands.} 
	\label{fig:epn_inpta_comparison}%
\end{figure}

Inspection of the EPN polarization profiles \citep{EPN_paper} reveals the presence of five identifiable features across the pulse window (see Fig.\ref{fig:epn_profiles}). Three of these features exhibit remarkably stable phase locations over a broad frequency range extending approximately from 728 MHz to 1369 MHz (see Fig.\ref{fig:epn_inpta_comparison}). The remaining two features, however, show modest but systematic variations in their phase locations with observing frequency. Such behavior is not unexpected. Within the framework of radius-to-frequency mapping, radiation at different frequencies originates at different heights in the pulsar magnetosphere, causing the observer's line of sight to intersect slightly different regions of the emission cone. Additional contributions from aberration and retardation effects, as well as changes in the dominant polarization mode, can further shift the apparent locations of intensity peaks and polarization features.

Although full-stokes observations are not available for the InPTA Band-3 data, the frequency-integrated profiles at 300--500 MHz continue to exhibit morphological signatures that are broadly consistent with the five emission regions inferred from the higher-frequency polarization data (Figure~\ref{fig:epn_inpta_comparison}), where we compare InPTA Band-3 profile aligned with the EPN profiles at 728 and 1369 MHz. In particular, the locations of three features (marked with grey dashed vertical lines in the Fig.\ref{fig:epn_inpta_comparison}), remain largely consistent with those seen at higher frequencies. However, the second feature in the linearly polarized profile (Fig.\ref{fig:epn_profiles}) exhibits a gradual shift toward the central emission region with decreasing frequency. Although full-Stokes observations are unavailable in the InPTA Band-3 range, the corresponding features cannot be independently identified from the total-intensity morphology at these frequencies. Additionally, one of the leading features appears at a slightly displaced phase location relative to its high-frequency counterpart. These behaviours suggest that, while the observed morphology is consistent with corresponding emission features persisting across frequency, the apparent locations of individual components need not remain strictly invariant. This further motivates the use of informative phase priors that permit modest excursions of the centroid of the Gaussian components rather than enforcing fixed phase locations.

The above considerations motivate our choice to model the pulse profile of PSR J1713+0747 using five Gaussian components. Rather than imposing rigid phase locations for these components, we employ informative priors centered on the phase locations inferred from the EPN polarization data while allowing modest excursions around these values. Such an approach preserves the physical association between Gaussian components and underlying emission regions while accommodating intrinsic profile evolution and frequency-dependent shifts in the apparent component locations. The implementation of these informative priors and the Bayesian decomposition framework is described in the following section.

\section{Bayesian Gaussian Decomposition}
\label{sec:bayesian}

\subsection{Gaussian profile model}

We model the pulse profile as the sum of 5 Gaussian components, $G_i$'s,

\begin{equation}
P(\phi)=\sum_{i=1}^{5}
G_i,~with~~
G_i=C_i
\exp\left[
-\frac{(\phi-\phi_i)^2}
{2\sigma_i^2}
\right],
\end{equation}

where $C_i$, $\phi_i$, and $\sigma_i$ denote the amplitude, centroid phase, and width of the $i$th Gaussian component, respectively, and $\phi$ represents the pulse phase. The choice of five components is motivated by the emission features identified from the polarization properties of PSR J1713+0747, as discussed in Section~\ref{sec:motivation_5gaussian}.

\subsection{Construction of morphology-driven priors}

The determination of suitable priors is an important aspect
of the decomposition procedure. The priors are informed by the emission features identified from the full-polarization EPN data together with the
frequency-integrated InPTA Band-3 profiles. The five Gaussian components are identified with the morphological features
of the pulse profile, namely the leading edge ($G_1$), the two central emission regions ($G_2$ and $G_3$), the post-peak shoulder or
hump ($G_4$), and the trailing edge ($G_5$). 

Since these Gaussian components are intended to track persistent morphological features associated with the emission regions identified in Section~\ref{sec:motivation_5gaussian}, whose geometric ordering is expected to remain unchanged over the observing baseline, we adopt a common set of centroid-phase priors for each component throughout the entire dataset. The centroid-phase priors encode
the expected locations of the underlying emission regions,
whereas the amplitude and width priors characterize the expected contribution of each emission region to the observed pulse-profile morphology. Rather than imposing fixed component phase locations, the prior ranges are chosen to be sufficiently broad while remaining exclusive so as to preserve their component identity and intended association with the corresponding emission features, and minimize arbitrary exchanges between neighboring Gaussians during the Bayesian inference. The only relaxation to the exclusivity of the priors is made for
the central components, $G_2$ and $G_3$ . As motivated by the polarization analysis in Section \ref{sec:motivation_5gaussian}, these two emission
regions occupy nearly the same phase range in the InPTA
Band-3 profiles and are therefore assigned overlapping
centroid-phase priors. Their amplitude and width priors are, however, chosen independently to reflect their distinct contributions to the observed pulse-profile morphology. This combination provides sufficient flexibility to model the closely spaced central emission features while maintaining a consistent decomposition across the observing baseline.
In
particular the constraint $\phi_1 < \phi_2,
\phi_3 < \phi_4 < \phi_5$, is imposed to
preserve the overall profile morphology.

The prior ranges adopted throughout the observing baseline are listed in Table \ref{tab:priors}. Uniform priors are adopted
for the component amplitudes and centroid phases, while log-uniform priors are employed for the component widths.
The posterior distributions are then obtained using the
Bayesian framework described below.

\begin{table}[!t]
\centering
\caption{
Morphology-driven prior ranges adopted for the five-component Gaussian decomposition.
A common set of priors is employed for all observations throughout the observing baseline.
Here $\mathcal{U}$ and $\log_{10}\mathcal{U}$ denote uniform and log-uniform prior distributions, respectively.
}
\label{tab:priors}
\renewcommand{\arraystretch}{1.15}

\begin{tabular}{lc}
\hline
Parameter & Prior \\
\hline

$C_1$      & $\mathcal{U}(0.03,\,0.16)$ \\
$C_2$      & $\mathcal{U}(0.40,\,0.80)$ \\
$C_3$      & $\mathcal{U}(0.20,\,0.40)$ \\
$C_4$      & $\mathcal{U}(0.15,\,0.70)$ \\
$C_5$      & $\mathcal{U}(0.04,\,0.25)$ \\

\hline

$\phi_1$   & $\mathcal{U}(0.40,\,0.46)$ \\
$\phi_2$   & $\mathcal{U}(0.47,\,0.52)$ \\
$\phi_3$   & $\mathcal{U}(0.47,\,0.52)$ \\
$\phi_4$   & $\mathcal{U}(0.53,\,0.58)$ \\
$\phi_5$   & $\mathcal{U}(0.59,\,0.65)$ \\

\hline

$\sigma_1$ & $\log_{10}\mathcal{U}(0.006,\,0.045)$ \\
$\sigma_2$ & $\log_{10}\mathcal{U}(0.010,\,0.040)$ \\
$\sigma_3$ & $\log_{10}\mathcal{U}(0.002,\,0.010)$ \\
$\sigma_4$ & $\log_{10}\mathcal{U}(0.008,\,0.070)$ \\
$\sigma_5$ & $\log_{10}\mathcal{U}(0.010,\,0.090)$ \\

\hline
\end{tabular}
\end{table}

\subsection{Bayesian inference framework}

Given a profile consisting of $N$ phase bins with measured
intensities
$\mathbf{d}=\{d_1,d_2,\dots,d_N\}$,
we assume that the noise in each phase bin is independent and
Gaussian. The noise level, $\sigma_{off}$, is estimated as the standard deviation of the off-pulse intensities and is used as the per-bin uncertainty in the likelihood function.

Under this assumption, the likelihood may be written as

\begin{equation}
\mathcal{L}
=
P(\mathbf{d}|\boldsymbol{\theta})
=
\prod_{k=1}^{N}
\frac{1}
{\sqrt{2\pi\sigma_{off}^2}}
\exp
\left[
-\frac{
\left(d_k-M(\phi_k|\boldsymbol{\theta})\right)^2
}
{2\sigma_{off}^2}
\right],
\end{equation}

where $M(\phi_k|\boldsymbol{\theta})$ is the model profile evaluated at the $k$th phase bin and

\begin{equation}
\boldsymbol{\theta}
=
\{
C_i,\phi_i,\sigma_i
\}_{i=1}^{5}
\end{equation}

denotes the set of model parameters.

The posterior distribution of the model parameters is obtained
through Bayes' theorem,

\begin{equation}
P(\boldsymbol{\theta}|\mathbf{d})
=
\frac{
P(\mathbf{d}|\boldsymbol{\theta})
P(\boldsymbol{\theta})
}
{
P(\mathbf{d})
},
\end{equation}

where $P(\boldsymbol{\theta})$ represents the morphology-driven priors and $P(\mathbf{d})$ is the Bayesian evidence.

The posterior distributions are sampled using the {\tt PyMultiNest} sampler~\citep{pymultinest_Feroz, pymultinest_Buchner}, from which the median and credible intervals of the model parameters are obtained.

\subsection{Validation of the decomposition}

The quality of the decomposition is assessed by subtracting the observed profile from the profile reconstructed using the posterior median parameter values of the five Gaussian components and examining the statistical properties of the resulting residuals.
\begin{figure}[!h]
	\centering \includegraphics[width=0.48\textwidth]{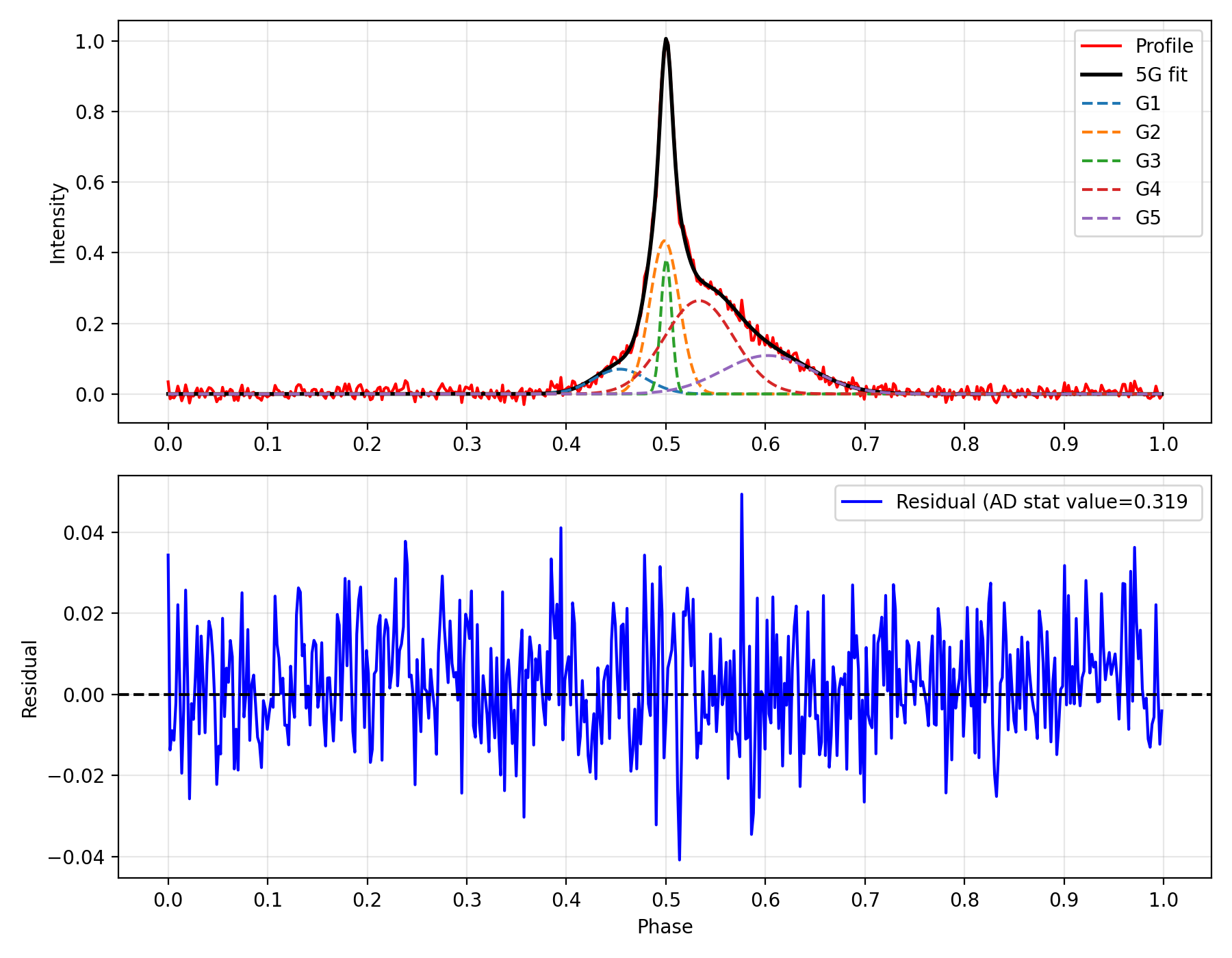}	
	\caption{Gaussian decomposition for a representative post-event epoch (MJD=59551). The upper panel shows the observed pulse profile (red), the individual Gaussian components (G1–G5; dashed curves), and their summed best-fit model (black). The lower panel shows the corresponding fit residuals, with the Anderson–Darling (AD) statistic indicated in the legend.} 
	\label{fig:example_bayesian_gaussian}
\end{figure}
The residuals are normalized by subtracting their mean and dividing by their standard deviation. The Anderson--Darling (AD) statistic of these normalized residuals is employed here as a goodness-of-fit diagnostic to compare the quality of the Gaussian decomposition across epochs. From our data, the AD critical values corresponding to the 5\% and 1\% significance levels are 0.752 and 1.076, respectively. These values provide convenient reference thresholds, with the latter serving as a more conservative one. AD statistics value below this threshold indicate residuals that are consistent with a Gaussian distribution, while progressively larger values indicate increasing departures from Gaussianity and may point to unmodelled profile structure. For completeness, we also report the $p$-value associated with the AD statistic, which expresses the significance of the observed departure from Gaussianity under the null hypothesis -- larger $p$-values indicate that the observed residuals are statistically consistent with a Gaussian distribution, whereas values below the conventional significance threshold of $p=0.05$ suggest a statistically significant departure from Gaussianity. As an example of our successful bayesian gaussian decomposition of a post-event epoch (MJD=59551) together with its associated AD statistic (see Fig.\ref{fig:example_bayesian_gaussian}) -- the corresponding $p$-value is 0.528,

The above modeling is, therefore, considered satisfactory when the residuals exhibit no significant departures from Gaussianity across all epochs.

\subsection{Application to frequency-resolved profiles}
\label{sec:freq_scrunch_analysis}

The profile-change event of PSR~J1713+0747 introduces a significant complication in the estimation of dispersion measure (DM) variations. Since the post-event profiles differ substantially from the template constructed using pre-event observations, the traditional determination of time-of-arrival (ToA) through cross-correlation, and hence the estimation of epoch-by-epoch DM variations, becomes unreliable. A common approach in such situations is to adopt a constant DM, typically chosen to be the median value measured during the stable pre-event baseline, and apply it throughout the entire dataset.

However, the DM of the pulsar is expected to vary over the long observational baseline. Adopting a constant DM can therefore leave residual dispersive delays unmodelled, leading to frequency-dependent variation of the integrated pulse profile. Such chromatic effects may then be inadvertently absorbed by the Gaussian decomposition and subsequently bias the timing analysis. An accurate determination of the DM at each observing epoch is therefore essential to disentangle propagation-induced chromatic effects from intrinsic profile evolution and to ensure that the subsequent profile-domain timing analysis remains free from such systematic biases.

\begin{figure}[!h]
	\centering \includegraphics[width=0.48\textwidth]{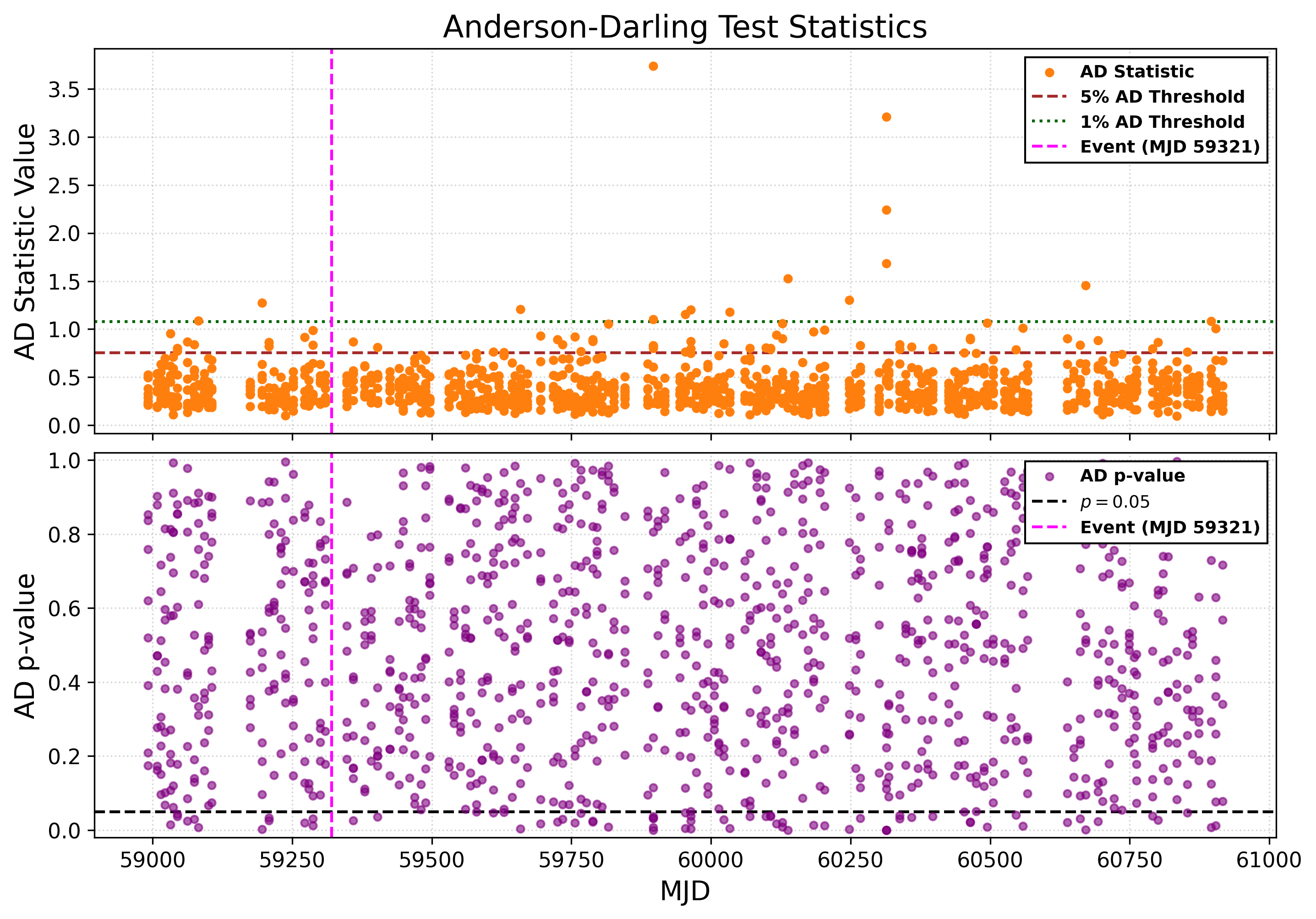}	
	\caption{The Gaussianity test (AD-test) of the sub-banded profile residuals across the observing baseline. The upper panel shows the Anderson–Darling (AD) statistic, with the horizontal dashed and dotted lines indicating the 5\% and 1\% critical values (AD = 0.752 and 1.076), respectively. The lower panel shows the corresponding AD p-values, with the horizontal dashed line $p$=0.05 serving as the conventional threshold value, above which the Gaussianity test succeeds. The magenta dashed vertical line in both panels marks the profile-change event at MJD 59321}
	\label{fig:AD_stat_subbanded}
\end{figure}

For this reason, the final Gaussian decomposition is performed on frequency-resolved profiles rather than on frequency-integrated profiles. The observing bandwidth is divided into eight sub-bands, providing sufficient frequency resolution to capture the profile evolution across the 300--500 MHz band while maintaining adequate signal-to-noise ratio within each sub-band. 
The methodology, described above, for the frequency-integrated analysis, is then applied to the profile in each sub-band. Figure~\ref{fig:AD_stat_subbanded} summarizes the AD statistics across the observing baseline, showing that the overwhelming majority of epochs lie below the 5\% significance threshold, with only a small number of isolated epochs exceeding the conservative 1\% criterion. The corresponding AD $p$-values are consistently greater than the conventional significance threshold of $p=0.05$, supporting the conclusion that the adopted Gaussian decomposition provides an adequate description of the observed pulse profiles across the frequency sub-bands and across epochs.

\section{Evolution of the Gaussian Components}
\label{sec:evolution_gauss}

The frequency-resolved Gaussian decomposition described in the previous section enables the temporal evolution of the individual Gaussian components associated with the emission features of PSR J1713+0747 to be tracked throughout the observing baseline. In particular, the inferred centroid phases provide a measure of the temporal localization of these components and their response to the profile-shape changes associated with the 2021 event.

\begin{figure}[!h]
\centering

\begin{subfigure}{0.98\columnwidth}
    \centering
    \includegraphics[width=\linewidth]{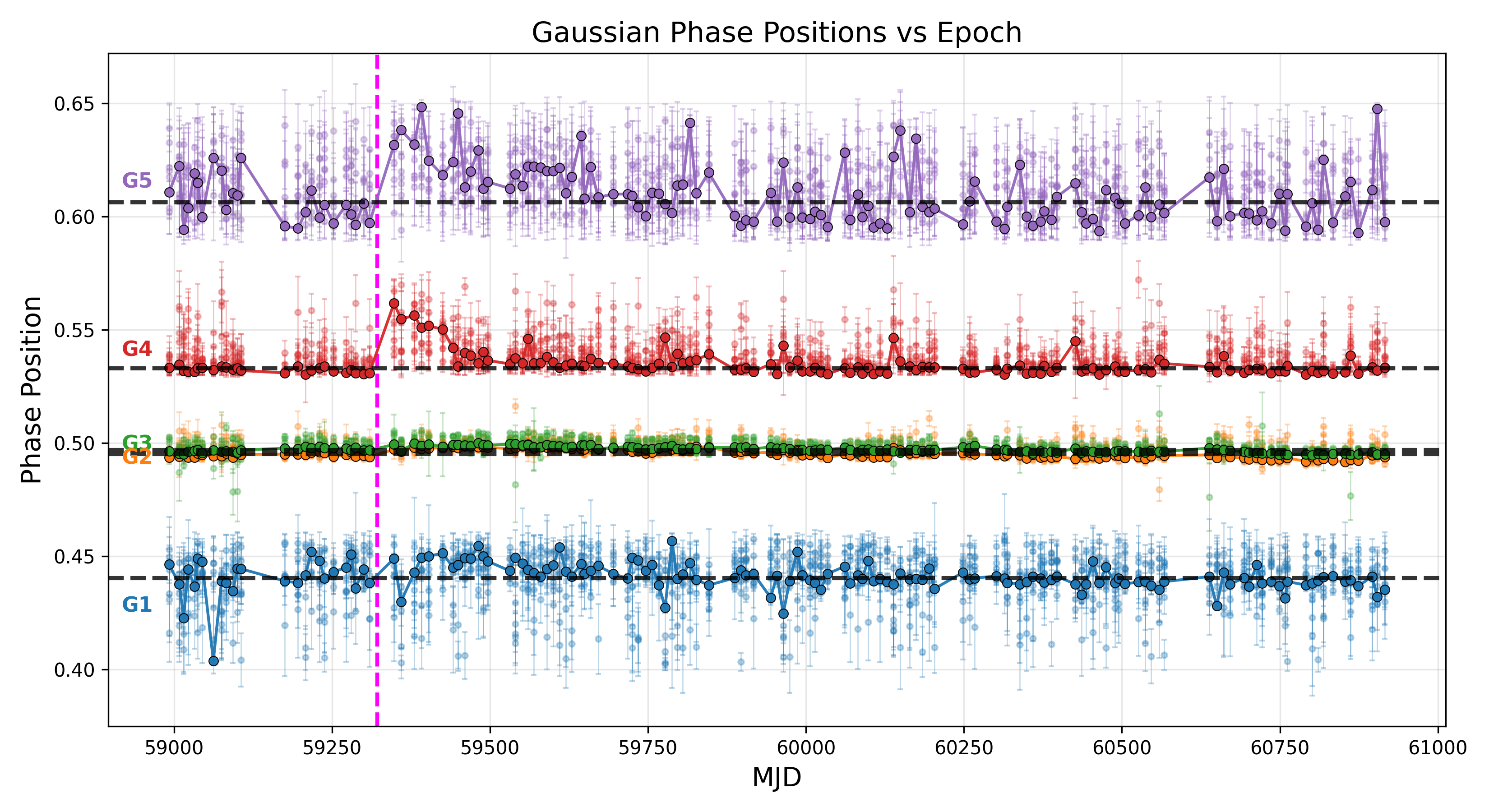}
    \caption{}
    \label{fig:5a}
\end{subfigure}

\vspace{0.2cm}

\begin{subfigure}{0.98\columnwidth}
    \centering
    \includegraphics[width=\linewidth]{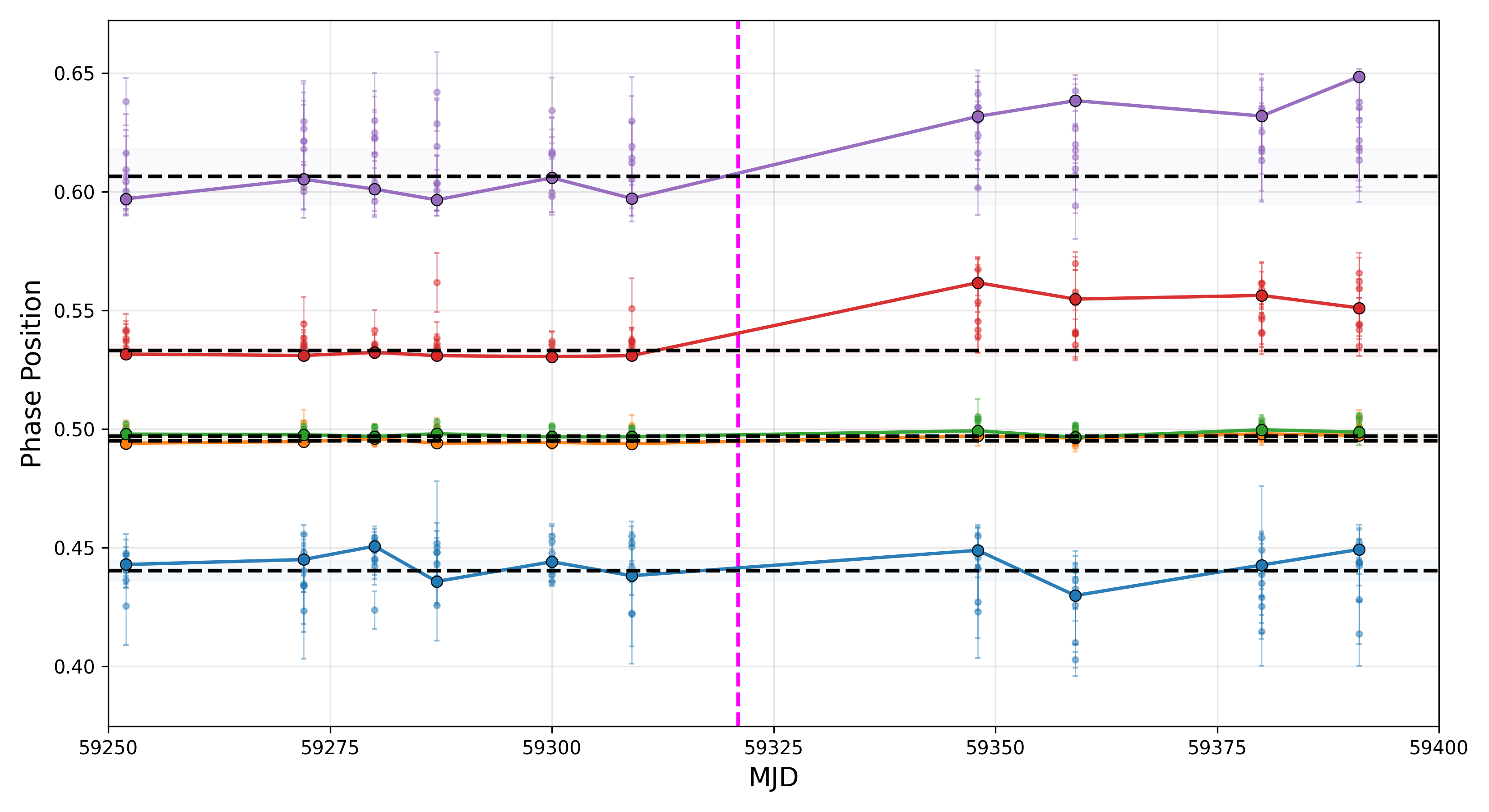}
    \caption{}
    \label{fig:5b}
\end{subfigure}
	
\caption{Phase evolution of the five Gaussian components corresponding to sub-banded profiles across the observing baseline. The individual points represent the posterior median phase values, with error bars showing symmetric uncertainties derived
from the central 68\% posterior credible intervals. The connected curves trace the weighted-median phase at each observing epoch. The magenta dashed vertical line marks the profile-change event at MJD 59321, while the black dashed horizontal lines indicate the full-baseline median phase of the corresponding Gaussian components. (a) Evolution over the full observing baseline. (b) Enlarged view around the profile-change event, highlighting the comparatively stable centroid phases of $G_2$ and $G_3$ and the larger phase excursions exhibited by the outer components.} 
	\label{fig:phase_evolution}
\end{figure}

Fig.~\ref{fig:5a} presents the evolution of the centroid phases of all five Gaussian components obtained from the sub-banded analysis. The horizontal dashed lines indicate the median centroid phase of each Gaussian component computed over the full observing baseline and serve as reference levels against which the inferred phase positions at each epoch can be compared. A clear hierarchy in the distribution of the recovered centroid phases is evident. The central components, $G_2$ and $G_3$, remain tightly clustered about their respective median phase positions throughout the observing span, including across the profile-change event (Fig.~\ref{fig:5b}), suggesting that the central emission region remains comparatively stable against the profile-shape changes. In contrast, the remaining components ($G_1$, $G_4$ and $G_5$) exhibit broader distributions of recovered centroid phases and generally larger posterior uncertainties. The recovered centroid phases, therefore, demonstrate that $G_2$ and $G_3$ are the most consistently localized components recovered by the Bayesian decomposition over the full observing baseline. Hereafter, we refer to $G_2$ and $G_3$ as the stable central components, where stability primarily denotes their persistent centroid-phase localization over the observing baseline.

In the conventional core-cone picture of pulsar emission, the central component is often referred to as the core component. It is expected to be stable, as it is hypothesised to originate from a region close to the stellar surface and to be less affected by radius-to-frequency mapping than the surrounding conal emission, since the associated magnetic field lines are much closer together \citep{Rankin1983}. Therefore, this central component can be assumed to be the closest to an invariant point on the stellar surface measuring the rotational phase. The preferential phase stability of $G_2$ and $G_3$ (Fig.\ref{fig:phase_evolution}), which model the central profile region, is consistent with their possible association with a stable core-like emission region, although the underlying emission geometry cannot be uniquely established. We, therefore, reconstruct the pulse profiles in the following section using only the stable central Gaussian components, $G_2$ and $G_3$, for subsequent dispersion-measure estimation and timing analysis.

\section{Reconstruction of Stable Profiles}
\label{sec:reconstruction}

The analysis presented in Section~\ref{sec:evolution_gauss} demonstrates that the central Gaussian components, $G_2$ and $G_3$, exhibit stability against the profile-shape changes observed in PSR~J1713+0747. These components are, therefore, expected to preserve
a substantial fraction of the timing information despite the
significant profile-shape variations exhibited by
PSR~J1713+0747.

\begin{figure}[!h]
\centering

\begin{subfigure}{0.86\columnwidth}
    \centering
        \includegraphics[width=\linewidth]{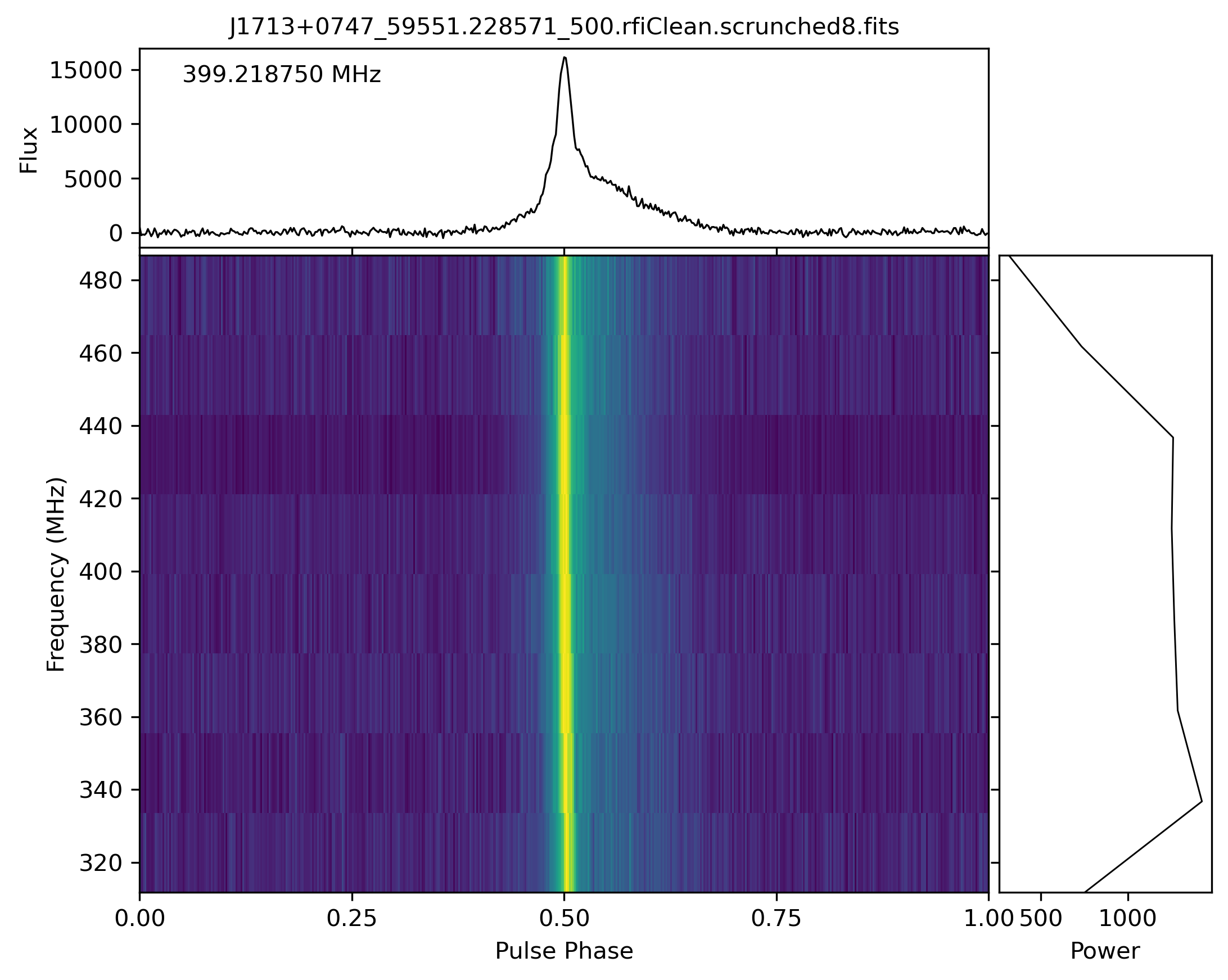}
    \caption{}
    \label{fig:example_gauss_decomp}
\end{subfigure}
\vspace{0.1cm}
\begin{subfigure}{0.86\columnwidth}
    \centering
    \includegraphics[width=\linewidth]{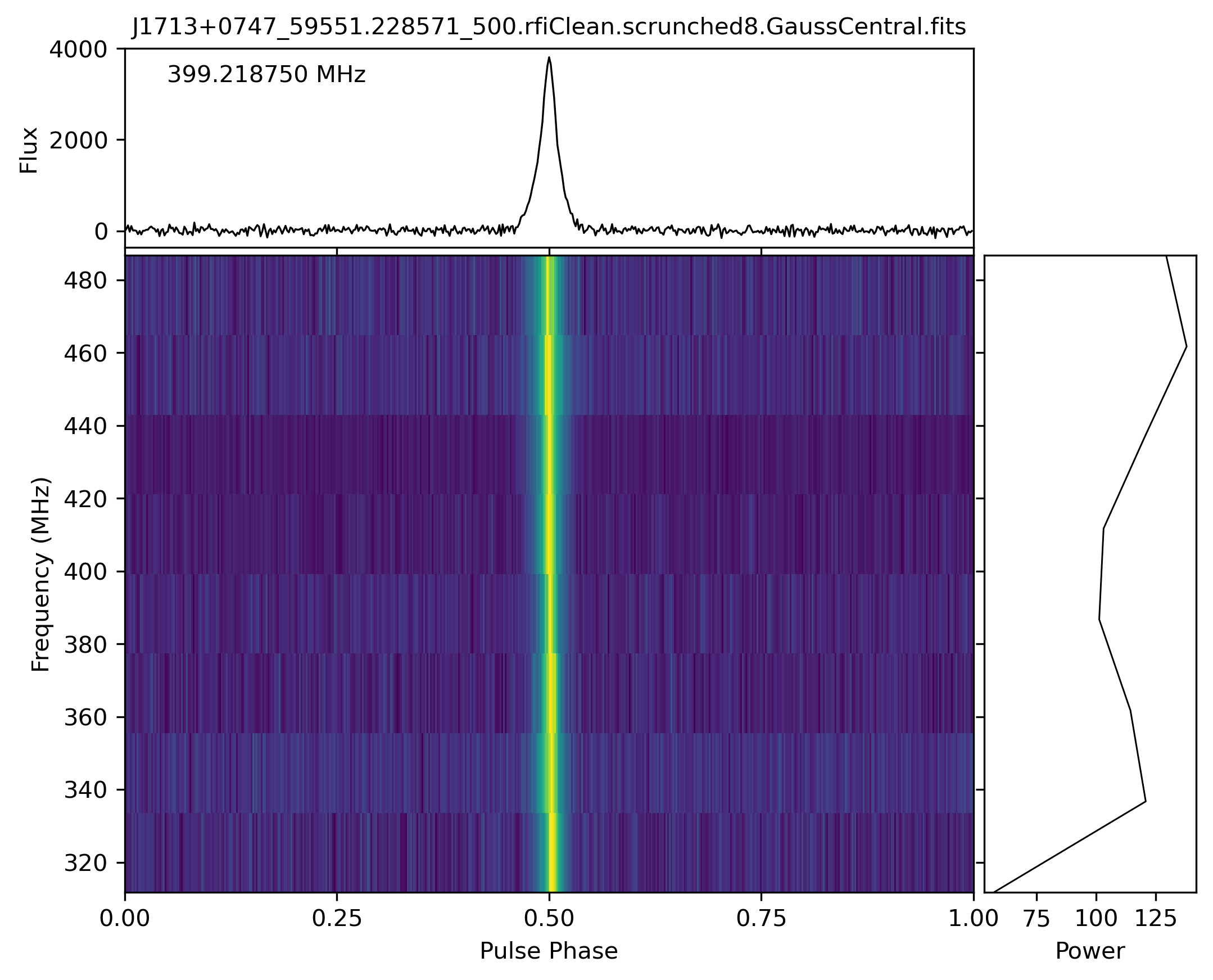}
    \caption{}
    \label{fig:example_posterior}
\end{subfigure}

\caption{Reconstruction of stable pulse profile from the original profile. For a representative post-event MJD 59551, we show (a) original frequency resolved profile for 8 sub-bands and (b) corresponding reconstructed profile with same frequency resolution, using stable central Gaussian components $G_2$ and $G_3$. }
\label{fig:original_reconstructed_profile}
\end{figure}

For each observing epoch and frequency sub-band, we reconstruct a profile using only the two stable central components,

\begin{equation}
P_{\rm rec}(\phi)=G_2(\phi)+G_3(\phi),
\label{eq:reconstructed_profile}
\end{equation}

where $G_2(\phi)$ and $G_3(\phi)$ are evaluated using the posterior median values of their corresponding Gaussian parameters obtained from the Bayesian analysis. In addition to the noise-free reconstructions, we generate noise-realistic profiles by adding a Gaussian noise realization with zero mean and standard deviation equal to the off-pulse rms of the corresponding peak-normalized profile. This preserves the characteristic noise level of the observations, while the noise-free profiles are retained for template construction. For each observing epoch and each of the eight frequency sub-bands, we, therefore, have two representations of the reconstructed profile: a noise-free profile and a noise-realistic profile. 

The reconstructed profiles are subsequently written back into the original frequency-resolved PSRFITS archives. Starting from the original archive of a given epoch, the profile amplitudes in each frequency channel are replaced by the reconstructed profile corresponding to that sub-band. In this manner, only the pulse-profile data are modified, while all metadata and header information associated with the original observations are preserved (see Fig.\ref{fig:original_reconstructed_profile}).

These reconstructed archives are subsequently used within the conventional pulsar timing framework for the estimation of epoch-wise dispersion measures, the generation of times of arrival, and the timing analysis presented in the following sections.

\section{Dispersion Measure}
\label{sec:dm_estimation}

The noise-free reconstructed profiles, generated using only the
stable central components $G_2$ and $G_3$, are used to construct
a frequency-resolved timing template. As a conservative choice, we adopt the reconstructed profile
corresponding to MJD~59300 as the reference template. This epoch
serves as the template reference in the InPTA-DR2 timing analysis,
providing a well-established fiducial reference for the present
work. The template
consists of eight frequency sub-bands, matching the frequency
resolution of the reconstructed archives.

\begin{figure}[!h]
	\centering \includegraphics[width=0.48\textwidth]{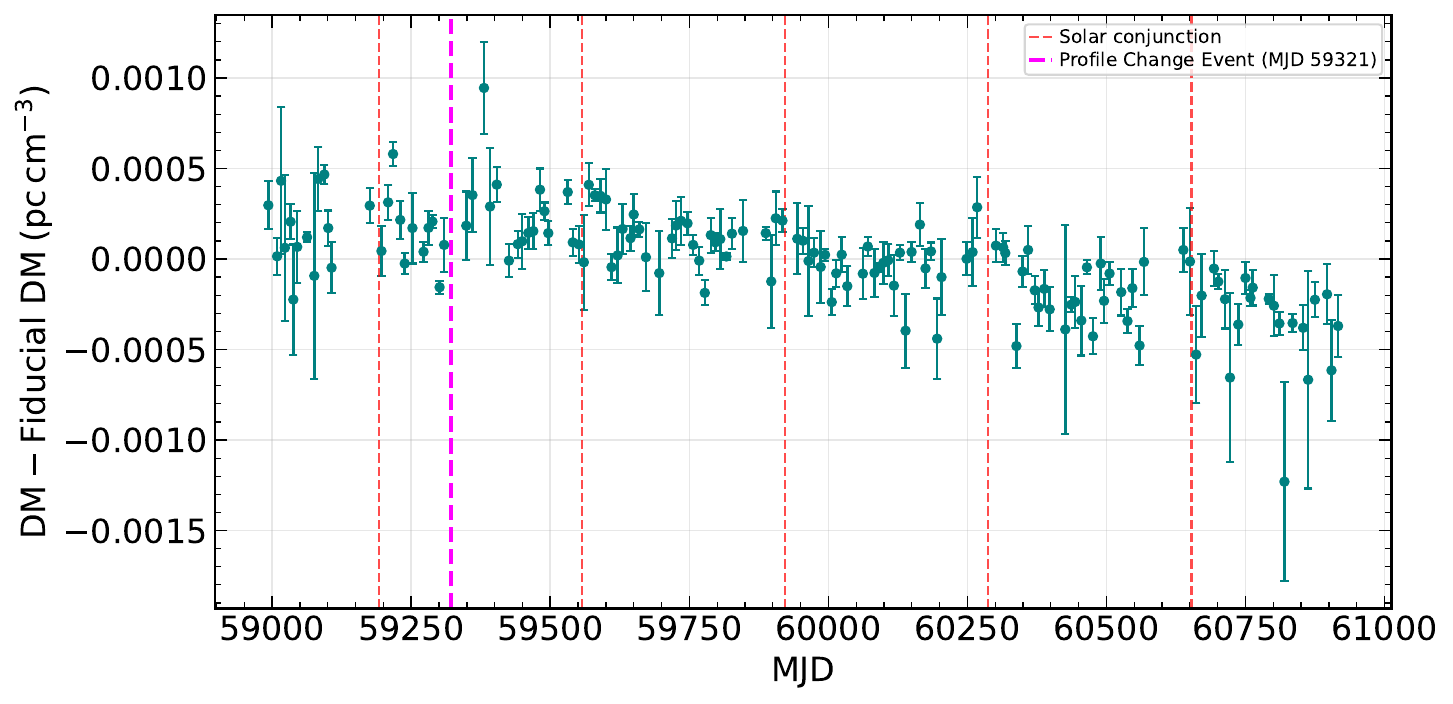}	
	\caption{DM variations of PSR J1713+0747 measured using DMCalc with the reconstructed profiles. Here we plot the difference w.r.t. the template DM (fiducial DM for reference) we started with. The red vertical dotted lines indicate annual solar conjunctions. The profile change event epoch is indicated with the magenta colored dashed lines.} 
	\label{fig:DMvsMJD}
\end{figure}

To estimate the dispersion measure (DM) at each observing
epoch, we employ the {\tt DMCalc} script
\citep{KK_DMCalc2021}. For a given epoch,
it cross-correlates each of the eight sub-band
profiles with the corresponding sub-bands of the frequency-resolved
template to generate sub-banded times of arrival (ToAs) and their
associated residuals. Following the standard statistical
filtering of unreliable or outlying ToAs, the frequency
dependence of the residuals is modelled according to the
dispersive delay relation, $t_{\rm DM} \propto \nu^{-2}$,
thereby yielding the best-fitting DM corresponding to that
epoch. The generated DM time series (Fig.~\ref{fig:DMvsMJD})
shows modest DM excursions in the vicinity of the annual solar
conjunctions, consistent with the expected contribution from
the solar wind. Since PSR~J1713+0747 has an ecliptic latitude of
approximately $30^\circ$, its minimum Sun--pulsar angular
separation lies close to the cut-off typically adopted in IPTA
timing analyses for avoiding significant solar-wind
contamination. The recovery of these expected annual
variations provides an empirical indication that the reconstructed profiles preserve
the chromatic information necessary for robust epoch-wise
DM estimation.

The resulting epoch-dependent DM estimates (see Fig.\ref{fig:DMvsMJD}) are subsequently used
to correct for the chromatic time-delays, induced by dispersion, in the ToA residuals. The associated timing analysis is presented in the following section.

\section{Timing}
\label{sec:timing}

Having reconstructed the frequency-resolved pulse profiles from
the stable central emission components and determined the
epoch-dependent dispersion measures, we now investigate whether the reconstructed profiles can be used for precision pulsar timing.

As an initial validation of the reconstruction procedure, we
compare the timing residuals obtained from the original and
reconstructed profiles during the pre-event interval. For this
comparison, both datasets are analyzed using the same EPTA
timing solution, which contains a single fiducial DM and no
epoch-dependent DMX corrections. The purpose of this exercise
is solely to assess whether the profile reconstruction preserves
the phase information required for conventional template
matching, independent of the DM estimation procedure described
in Section~\ref{sec:dm_estimation}.

\begin{figure}[!h]
	\centering \includegraphics[width=0.48\textwidth]{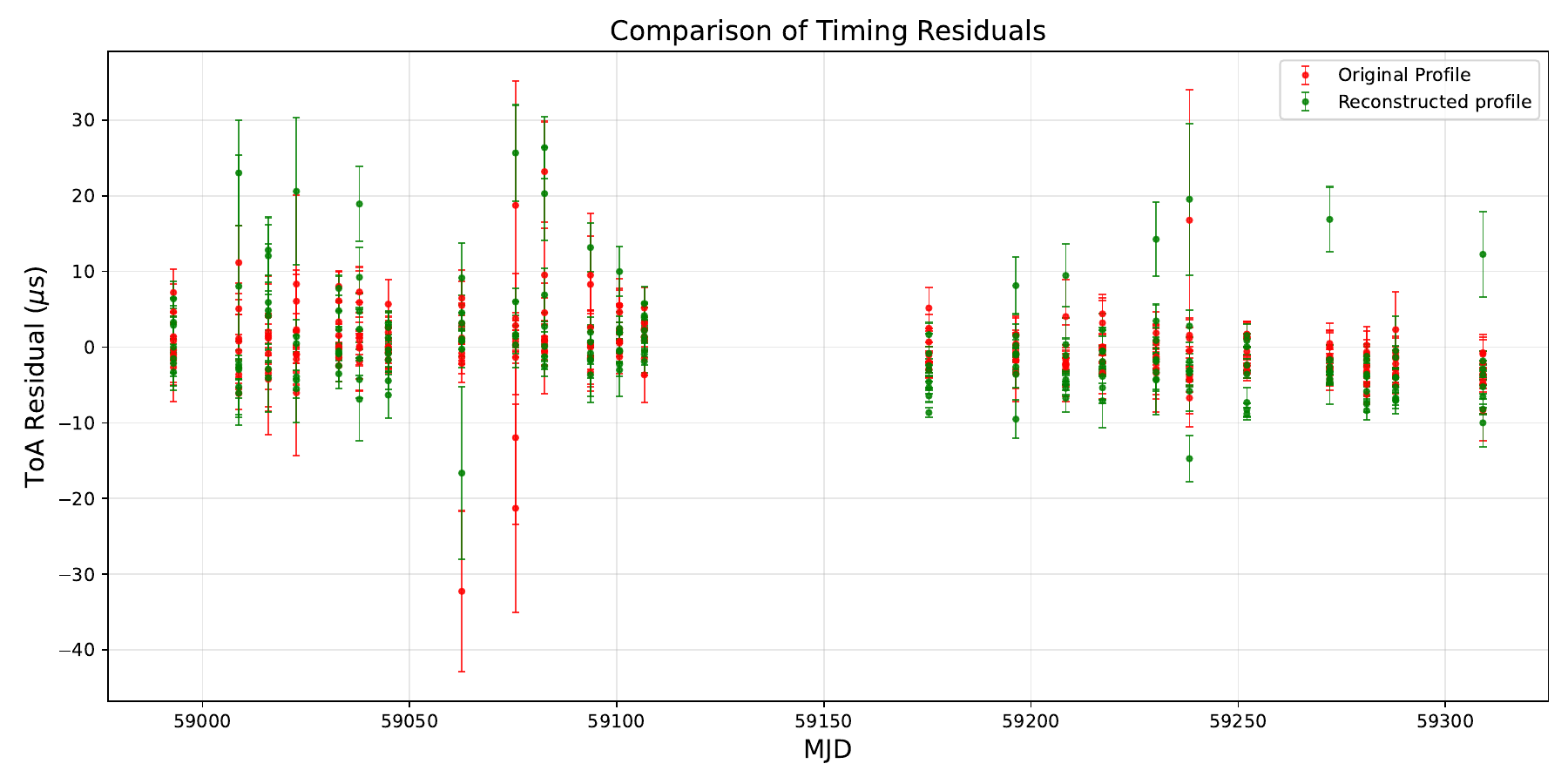}	
	\caption{The ToA residuals, from pre-event data, obtained from the full profile morphology (red color) and the ones obtained from the reconstructed profile (green color) from the central gaussian components.} 
	\label{fig:pre_event_timing_compare}
\end{figure}

Figure~\ref{fig:pre_event_timing_compare} compares the timing residuals
obtained from the original pulse profiles with those derived
from the reconstructed profiles. The reconstructed profiles
recover the same overall temporal behavior of the residuals,
with no discernible systematic offsets relative to the original
data. The corresponding weighted root-mean-square (WRMS)
residuals are $2.505~\mu$s and $3.707~\mu$s for the original
and reconstructed profiles, respectively.
Furthermore, the original and reconstructed analyses do not
necessarily yield identical sets of sub-banded ToAs owing to
their independent quality filtering and channel flagging.
Consequently, this comparison is intended as a validation that
the reconstructed profiles preserve the timing information
contained within the stable emission region, rather than as a
one-to-one comparison of individual ToAs.

The above comparison is restricted to the pre-event interval,
where the pulse morphology remains well represented by the
reference template and conventional template matching is
expected to produce reliable ToAs. Following the profile-change
event, however, the substantial mismatch between the observed
profiles and the reference template compromises the reliability
of conventional ToA and DM estimation. Consequently, the
post-event timing analysis is performed exclusively using the
reconstructed profiles.

The final timing analysis is carried out using the reconstructed
frequency-resolved archives together with the epoch-dependent
DM values obtained from the procedure described in
Section~\ref{sec:dm_estimation}. 

\begin{figure}[!h]
	\centering \includegraphics[width=0.48\textwidth]{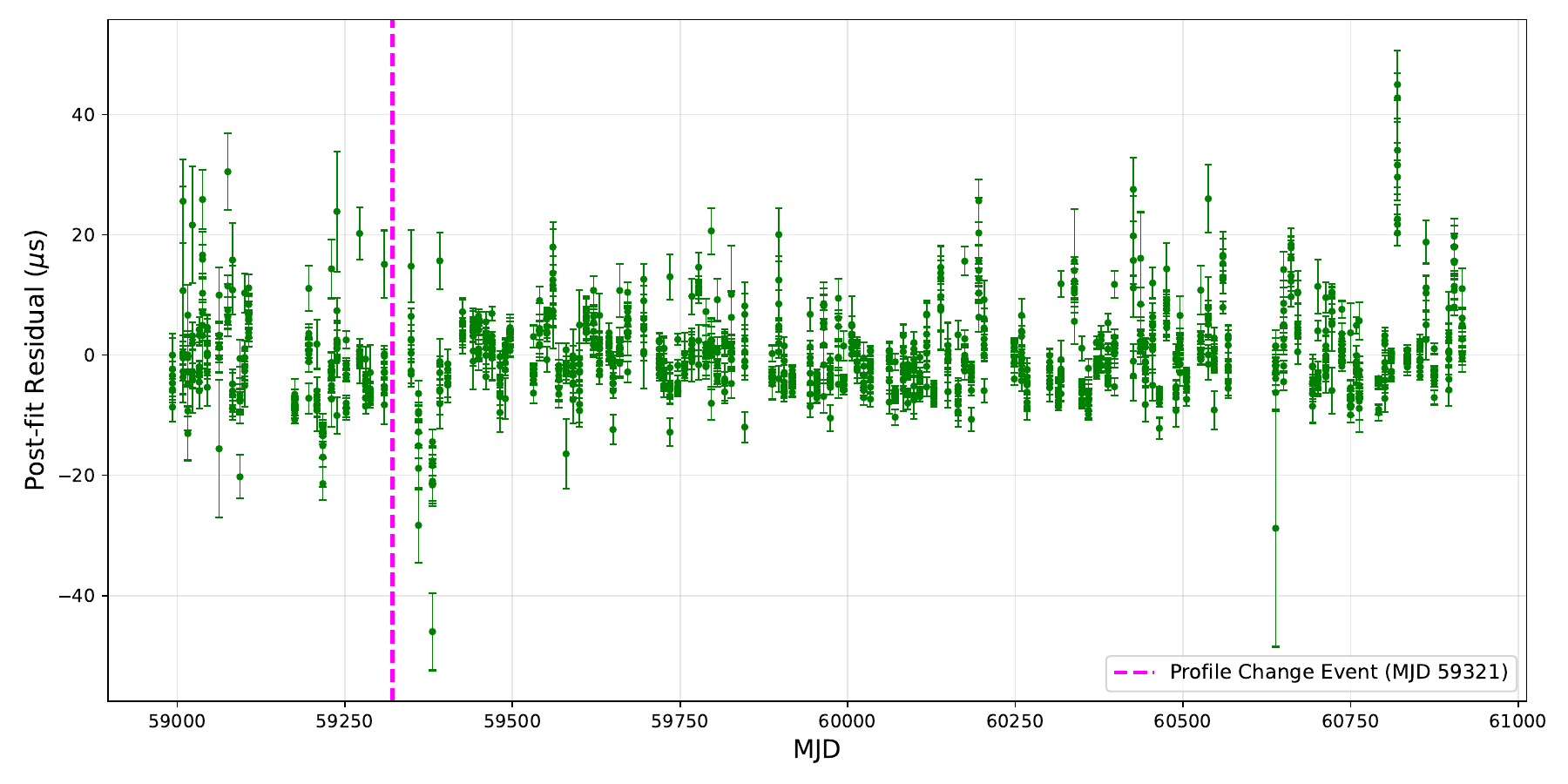}	
	\caption{Post-fit Timing residuals obtained using reconstructed profiles across the full observing baseline. The magenta vertical line marks the profile change event. The wrms obtained is $4.454~\mu$s.} 
	\label{fig:finaltiming}
\end{figure}

The ToAs, thus obtained, are analyzed using the standard
\texttt{tempo2}~\citep{TEMPO2} timing procedure. We find that fitting for the spin frequency F0 yields a stable timing solution, while fitting additional timing-model parameters does not lead to any statistically significant improvement in the post-fit residuals, indicating that the reconstructed data are adequately described. The resulting timing solution remains phase connected throughout the entire baseline, including the profile shape change event, and yields a
post-fit weighted root-mean-square residual of
$4.454~\mu$s (Fig.~\ref{fig:finaltiming}).

The successful recovery of a stable timing solution using
profiles reconstructed solely from the two temporally stable
Gaussian components demonstrates that the essential timing
information is preserved within the central emission region of
PSR~J1713+0747. More importantly, it shows that isolating these
stable emission components substantially mitigates the impact of
the long-term profile evolution on conventional timing,
providing a physically motivated route towards precision timing
of pulsars exhibiting abrupt pulse-profile changes and subsequent long-term profile evolution.

To ensure that the effect of the profile shape change event on the TOAs are corrected effectively by our procedure, we performed Bayesian analysis of the obtained TOAs to check for the presence of an exponential dip signature as done in \citet{Jaikhomba2021}\footnote{This analysis is done using the DISCOVERY package \citep{Vallisneri_nanograv_discovery_2025} in combination with Hamiltonian Monte Carlo implemented in the NUMPYRO package \citep{2019arXiv191211554P}.}.
We find no evidence for the exponential dip signal in our TOAs, unlike in \citet{Jaikhomba2021}, consistent with our reconstruction substantially mitigating the timing signature associated with the profile-change event.

\section{Discussion and Conclusions}
\label{sec:conclusion}

In this work we have presented a physically motivated
profile-domain framework for mitigating the effects of pulse
profile variability on precision pulsar timing. Rather than
treating the pulse profile as a single evolving template, the
method decomposes each frequency-resolved profile into
Gaussian components associated with persistent emission
regions through morphology-driven Bayesian priors. Tracking
these components across the observing baseline enables the
identification of emission features whose associated Gaussian components remain phase stable despite the significant profile-shape evolution exhibited by PSR~J1713+0747.

The analysis reveals that the two central Gaussian
components preserve remarkably stable
centroid phases throughout the observing baseline, whereas
the outer components exhibit substantially larger temporal
variations. This behavior suggests that the central emission
region retains the most reliable rotational phase information
even during the profile-change event. Although the present
analysis does not uniquely determine the underlying emission
geometry, it demonstrates that identifying stable emission
regions directly from the observed pulse morphology provides
a robust basis for reconstructing stable pulse profiles for precision timing. 

The Gaussian decomposition is performed on frequency-resolved pulse profiles, thereby preserving the frequency-dependent information required for reliable epoch-wise dispersion-measure estimation while simultaneously isolating intrinsic profile evolution from the timing analysis. This capability is particularly valuable for the low-frequency InPTA uGMRT Band-3 (300--500 MHz) observations considered here, where chromatic propagation effects are more pronounced. The recovery of the expected annual DM
variations associated with solar conjunctions, together with
the phase-connected timing solution obtained across the
profile-change event, demonstrates that the reconstructed
profiles preserve both the chromatic propagation information
and the rotational phase information required for
high-precision timing.

The reconstructed profiles reproduce the pre-event timing
behavior of the original observations while enabling a stable
timing solution throughout the full observing baseline. The
final timing solution remains phase connected across the
profile-change event and yields a post-fit weighted
root-mean-square residual of $4.454~\mu$s. This
demonstrates that a significant fraction of the timing
information is retained within a subset of the pulse profile,
even when the overall profile morphology undergoes
substantial evolution.

Although the methodology has been developed using
PSR~J1713+0747 as a representative test case, it is not
specific to this pulsar. The Bayesian framework is readily
adaptable to other pulsars exhibiting abrupt profile changes,
gradual secular evolution, mode changing, or other forms of
long-term pulse-shape variability provided high quality polarization data 
is available for relevant observing frequencies. The morphology-driven
priors may likewise be modified to accommodate different
profile complexities and emission geometries. As increasingly
sensitive PTA observations become available with instruments
such as the uGMRT, MeerKAT, FAST, and the
Square Kilometre Array (SKA), physically motivated profile-domain
approaches of this kind are expected to become increasingly
important for preserving the full timing potential of pulsars
whose profiles evolve with time.

The present work therefore provides a general framework for
identifying stable emission regions, reconstructing robust
pulse profiles, and recovering reliable timing information in
the presence of pulse-profile evolution. We anticipate that
this methodology will complement existing profile-domain
timing techniques and provide a practical route towards
maintaining high-precision timing for variable pulsars in
future PTA experiments.

\section*{Acknowledgements}
SD is supported by ANRF MTR\_2023\_000384. 
The work of CD at the Physical Research Laboratory (PRL) was supported by the Department of Space, Government of India. 
BCJ acknowledges the support from Raja Ramanna Chair fellowship of the Department of Atomic Energy, Government of India (RRC – Track I Grant 3/3401 Atomic Energy Research 00 004 Research and Development 27 02 31 1002//2/2023/RRC/R\&D-II/13886 and 
1002/2/2023/RRC/R\&D-II/14369). 
JS acknowledges the support from the University of Cape Town Vice Chancellor’s Future Leaders 2030 Awards programme and the South African Research Chairs Initiative of the Department of Science and Technology and the National Research Foundation.
KT is partially supported by JSPS KAKENHI grant Nos. 24H01813, 25K21670, 26H00838 and 26K21724.
VS acknowledges the support of the Department of Atomic Energy, Government of India, under project identification No. RTI 4002. 
KR is Funded by CSIR-NET (UGC JRF).
AKP is supported by CSIR fellowship Grant number 09/0079(15784)/2022-EMR-I.
ZZ is supported by the Prime Minister’s Research Fellows (PMRF) scheme, Ref. No. TF/PMRF22-7307.
S.J.D. was supported by IBS under the project code IBS-R018-D1.
HT is supported by DST INSPIRE Fellowship.
Supported by CSIR JRF Fellowship , Grant number 09/1020(20166)/2024-EMR-I.
NDB is supported by DST-WISE fellowship (DST/WISE-PDF/PM-17/2024).
Part of this research has made use of the EPN Database of Pulsar Profiles maintained by the University of Manchester, available at: \url{http://www.jodrellbank.manchester.ac.uk/research/pulsar/Resources/epn/.}

\appendix


\bibliographystyle{elsarticle-harv} 
\bibliography{main}






\end{document}